\documentclass[aps,twocolumn,showpacs,preprintnumbers,amsmath,amssymb,footinbib]{revtex4}
\usepackage[colorlinks=true, pdfstartview=FitV, linkcolor=red, citecolor=blue, urlcolor=green]{hyperref}
\newcommand{\alaalb}{\widehat{\alpha}^a \widehat{\alpha}^b}
\newcommand{\beq}{\begin{equation}}

\newcommand{\boldl}{{\bf L}}
\newcommand{\cala}{{\cal A}}
\newcommand{\call}{{\cal L}}
\newcommand{\calla}{{\cal L}_0}
\newcommand{\callb}{{\cal L}_1}
\newcommand{\callc}{{\cal L}_2}
\newcommand{\calm}{{\cal M}}

\newcommand{\calrn}{{\cal R}_{0}}
\newcommand{\cals}{{\cal S}}
\newcommand{\calseff}{{\cal S}^{\rm eff}}

\newcommand{\calv}{{\cal V}}

\newcommand{\calz}{{\cal Z}}
\newcommand{\cosalh}{\cos\left(\frac{\alpha}{2}\right)}
\newcommand{\cossqalh}{\cos^2\left(\frac{\alpha}{2}\right)}
\newcommand{\delab}{\delta^{a b}}
\newcommand{\eeq}{\end{equation}}
\newcommand{\ellad}{\ell_{ad}}

\newcommand{\fralt}{\frac{\alpha}{2}}
\newcommand{\overk}{k}
\newcommand{\overp}{p}
\newcommand{\real}{{\rm Re}}
\newcommand{\sinalh}{\sin\left(\frac{\alpha}{2}\right)}
\newcommand{\sinsqalh}{\sin^2\left(\frac{\alpha}{2}\right)}
\newcommand{\tillam}{\widetilde{\lambda}}
\newcommand{\tilxi}{\widetilde{\xi}}
\newcommand{\tr}{{\rm tr}\;}

\newcommand{\vebesq}{\vec{\beta}^{\,2}}
\newcommand{\vecal}{\vec{\alpha}}
\newcommand{\vecbe}{\vec{\beta}}
\newcommand{\veclepar}{\stackrel{\leftarrow}{\partial}}
\newcommand{\vecripar}{\stackrel{\rightarrow}{\partial}}
\newcommand{\vex}{{\vec x}}

\newcommand{\wha}{\widehat{\alpha}}

\newcommand{\zn}{$Z(N)\;$}
\def\anp#1#2#3{Annals Phys. {\bf #1}, #2 (#3)}
\def\atmp#1#2#3{Adv. Theor. Math. Phys. {\bf #1}, #2 (#3)}

\def\ibid#1#2#3{{\it ibid.} {\bf #1}, #2 (#3)}

\def\jhep#1#2#3{J. High Energy Phys. #2 (#3) #1}

\def\jpg#1#2#3{Jour. Phys. G {\bf #1}, #2 (#3)}
\def\npa#1#2#3{Nucl. Phys. A {\bf #1}, #2 (#3)}
\def\npb#1#2#3{Nucl. Phys. B {\bf #1}, #2 (#3)}
\def\npsb#1#2#3{Nucl. Phys. Proc. Suppl. B {\bf #1}, #2 (#3)}
\def\plb#1#2#3{Phys. Lett. B {\bf #1}, #2 (#3)}
\def\prc#1#2#3{Phys. Rev. C {\bf #1}, #2 (#3)}
\def\prd#1#2#3{Phys. Rev. D {\bf #1}, #2 (#3)}
\def\prl#1#2#3{Phys. Rev. Lett. {\bf #1}, #2 (#3)}
\def\phr#1#2#3{Phys. Rep. {\bf #1}, #2 (#3)}

\def\ptps#1#2#3{Prog. Theor. Phys. Suppl. {\bf #1}, #2 (#3)}
\def\rpp#1#2#3{Rep. Prog. Phys. {\bf #1}, #2 (#3)}
\def\rmp#1#2#3{Rev. Mod. Phys. {\bf #1}, #2 (#3)}

\def\zpc#1#2#3{Z. Phys. C {\bf #1}, #2 (#3)}
\begin{document}
\title{$\beta$-functions for a
$SU(2)$ Matrix Model in $2+\epsilon$ Dimensions}
\author{Michaela Oswald$^{a}$ and Robert D. Pisarski$^{b}$}
\affiliation{
$^a$Department of Physics,
University of Virginia, Charlottesville, VA, 22904, U.S.A.\\
$^b$Dept. of Physics,
Brookhaven National Lab., Upton, NY, 11973, U.S.A.\\
}
\begin{abstract}
To investigate the non-perturbative, electric sector
of a deconfined gauge 
theory at nonzero temperature, we consider a $SU(2)$ matrix model.
We compute $\beta$-functions to one loop order for the simplest
extension of the $O(4)$ nonlinear sigma model, which involves
three coupling constants.  Computing 
in the ultraviolet limit in $2 + \epsilon$ dimensions, we
find that at least one coupling is not asymptotically free.
\end{abstract}
\date{\today}
\maketitle

%%%%%%%%%%%%
\section{Introduction}
%%%%%%%%%%%%

The deconfining phase transition in $SU(N)$ gauge theories is of
interest in its own right, as a problem 
in statistical mechanics, and for its
possible application to the collisions of large nuclei at very
high energies.  

If the deconfining transition is of second order, then 
universality and the 
renormalization group predicts the critical
behavior \cite{svetitsky_yaffe}.  Lattice simulations indicate
that the deconfining transition is of second order 
for $N=2$, $N=3$, and (perhaps) $N=4$ in $2+1$ dimensions 
\cite{3Dlattice}, and for $N=2$ in $3+1$ dimensions \cite{su2}.

For other $N$, 
$N \geq 5$ in $2+1$ dimensions \cite{3Dlattice}, and for
$N \geq 3$ in $3+1$ dimensions \cite{su3,teper,bringoltz,holland},
the deconfining transition appears to be of first order 
\cite{su3,teper,bringoltz,holland}.
Universality is of no help for a first order transition.
Moreover, even when a transition
is of second order, universality is only of use 
near the transition; it says nothing about
what happens away from it.

A first order transition is natural at large $N$, 
because the number of degrees of freedom is $\sim 1$
in the confined phase, and $\sim N^2$ in the deconfined phase 
\cite{holland}.
One possibility is
that the transition is so strongly first order that 
the entire deconfined phase is a nearly ideal gluon plasma.
In $3+1$ dimensions, for example,
this might have been true for all $N$, from $N =3$ to $N=\infty$.
In this case, the deconfined phase would be amenable to analysis by means of a
resummed perturbation theory \cite{resum}, for all temperatures from
the transition temperature, on up.  
This does not seem to be what happens, however.
If $T_d$ is the temperature for deconfinement,
{\it all} consistent resummations of perturbation theory appear to fail
in describing temperatures below temperatures of order 
$\approx 4 T_d$ \cite{resum}.

A plausible guess is that at temperatures $T$ between
$T_d$ and $\approx 4 T_d$,
the theory goes into a regime of strong coupling \cite{peshier,shuryak}.  
After all, by running
asymptotic freedom in reverse,
as the temperature decreases, the value of
the strong coupling constant, $\alpha_s(T)$, increases.

Even at temperatures as low as the transition temperature, though,
the gauge coupling is not especially large.
Consider the effective
coupling which enters into the dimensionally reduced theory
in three dimensions \cite{helsinki}.  
Computations show that in 
the magnetic sector, nothing surprising happens, even at temperatures
as low as $T_d$.  For example, one can compare the static string tension in
the dimensionally reduced theory, to that in the full theory.
Using the two loop calculations of the magnetic coupling by
Giovannangeli \cite{coupling1}, Laine and Schr\"oder
\cite{coupling2} find remarkably good agreement between these
two quantities from high temperature all of the way down to $T_d$.  
They estimate that in QCD, even at a ``transition'' temperature
of $\approx 175$~MeV, that the effective coupling in the dimensionally
reduced theory is 
$\alpha^{\it eff}_s \approx 0.28$ \cite{coupling2}.  This is a surprisingly
small coupling: at zero temperature, it corresponds to a relatively high
momentum scale, of $\approx 1.6$~GeV.

While nothing remarkable happens in the magnetic sector, something
striking happens in the electric.  At nonzero temperature, 
the electric sector of a gauge theory is probed by 
the eigenvalues of the thermal Wilson
line, which are gauge invariant 
\cite{eigenvalues,derivative_terms,loop1,loop2,loop3,bielefeld,dhlop,aharony,dlp,rdp}.
Most notably, the Polyakov line
is proportional to the sum of these eigenvalues, and is near one 
in a perturbative regime.
Numerical simulations on the lattice show that while the 
renormalized Polyakov loop
is $\approx 0.9$ 
at $\approx 3 T_d$, it falls sharply as the temperature decreases,
to $\approx 0.4$ at the transition,
and vanishes in the confined phase \cite{bielefeld,dhlop}.  

Thus when a gauge theory deconfines, it does not go immediately to
a nearly ideal plasma at $T_d$.  Instead, for temperatures
between $T_d$, and $\approx 4 T_d$, the electric sector, although
not strongly coupled, behaves in a pronounced non-perturbative manner.

We remark that this behavior in the electric sector drives the theory
{\it far} from the conformal limit.  This is shown by 
the interaction measure, which is the trace
of the energy momentum tensor divided by $T^4$.  The interaction
measure vanishes for a conformally invariant theory, such
as for a free, massless field, or less trivially, for
a gauge theory with ${\cal N}=4$ supersymmetry.  
For an $SU(N)$ gauge theory in $3+1$ dimensions,
lattice simulations show that 
for $N=2$ \cite{su2}, $N=3$ \cite{su3}, $N=4$ \cite{bringoltz}, 
and $N=8$ \cite{bringoltz}, 
the interaction measure is very large when $T: T_d \rightarrow 4 T_d$.
How large can be estimated by comparing to the
value in perturbation theory, where it starts at 
$\sim \alpha_s^2$.  For $N=3$, the interaction measure has a sharp
maximum just above the transition, at $T \approx 1.2 T_d$.
At this maximum, the interaction measure is about
ten times larger than its perturbative value,
using the values of $\alpha_s(T)$ from \cite{coupling2}.
In contrast, by $\approx 4 T_d$, its 
value is within the range expected from perturbation theory.

In order to understand the 
non-perturbative, electric sector of a deconfined gauge theory, it
is necessary to develop an effective theory of 
thermal Wilson lines \cite{loop1,dhlop,aharony,dlp,rdp}.
Wilson lines are $SU(N)$ matrices, and so the effective theories
of relevance are matrix models.  

In this paper we take a small step in this process, by
considering the effective theory for deconfinement in $2+1$ dimensions.  
In $3+1$ dimensions, 
at least at large $N$
the effective theory for deconfinement is dominated by
an effective potential \cite{dlp}.
In $2+1$ dimensions, however, fluctuations
can dominate the potential.  Fluctuations are 
why, for instance, the deconfining transition for $N=3$ is of second
order in $2+1$ dimensions, 
versus the first order transition expected in mean field theory.

The infrared
behavior of such a matrix model is involved, and involves complex
interactions between the potential and fluctuations.  The ultraviolet
limit, though, is amenable to perturbative analysis, and it is this
which we study in the present work.

Our computation is similar to that for an ordinary nonlinear sigma
model in two dimensions 
\cite{zinn,rdp_symmetric,mckane_stone,friedan,principal_chiral,symmetric}.
While familiar sigma models have one coupling constant,
matrix models have an infinity of couplings, all of which are relevant in
the ultraviolet limit.  Thus
one of the purposes of this paper is to see if such models
are well behaved in perturbation theory.
We find that they are: modulo a reasonable technical assumption on the
size of the coupling constants, (\ref{coupling_restriction}),
we find that at least to one loop order,
the counterterms induced can be reabsorbed into
renormalizations of the bare coupling constants.  
This allows $\beta$-functions to be defined in a standard fashion
\cite{zinn}.

It is well known that the
single coupling constant of the nonlinear sigma model is 
asymptotically free in two dimensions 
\cite{zinn,rdp_symmetric,mckane_stone,friedan,principal_chiral,symmetric}.
One might hope that the new couplings in matrix models are
also asymptotically free.  We find that 
this is not true.  We consider the simplest
generalization of an $SU(2) \times SU(2)$ 
sigma model, which involves three coupling constants.
While the coupling analogous to that of the usual sigma model is always
asymptotically free, at least one of the two new couplings is not.  
This result is similar to that of Friedan, who considered
a sigma model in a background metric, and found that 
the new couplings induced by the background metric
usually spoil asymptotic freedom \cite{friedan}.  

In Sec. \ref{variety} we classify the types of matrix models
for general $N$; in Sec. \ref{sectiontwo}, for $N=2$.
Sec. \ref{bfm} describes the one loop calculation in a general 
background field.
The counterterms are computed in Sec. \ref{one_loop},
producing the $\beta$-functions of Sec. \ref{beta_functions_section}.
In Sec. \ref{conclusions} we suggest how the $\beta$-functions for $N=2$
might generalize to $N \geq 3$, and how these affect the
phase transitions of the model in, and above, two dimensions.
An appendix contains a comment about a nearly Goldstone boson at large $N$
in $SU(N)$ gauge theories.

%%%%%%%%%%%%
\section{Variety of Matrix Models}
\label{variety}
%%%%%%%%%%%%

Consider a matrix valued field $\boldl(\vec{x})$.  Let it be a
$SU(N)$ matrix, satisfying the constraints
\beq
\boldl^\dagger \boldl = {\bold 1} \;\;\; , \;\;\;
\det(\boldl) = 1 \; .
\eeq
We chose $\boldl$ to lie in the fundamental representation,
so $\boldl = \boldl_{a b}$, where $a,b = 1\ldots N$.
In the usual nonlinear sigma model,
this matrix is assumed to be invariant 
under a global symmetry of $(SU(N) \times SU(N))/Z(N)$,
\beq
\boldl(\vex) \rightarrow U \; \boldl(\vex) \; V \; ;
\eeq
$U$ and $V$ are independent global $SU(N)$ rotations,
modulo a common $Z(N)$ rotation.
If $U$ and $V$ are distinct rotations,
the only renormalizable Lagrangian invariant under this symmetry is 
\beq
\frac{1}{g^2} \; \tr \left| \partial_i \boldl \right|^2 \; ,
\label{nonlinear}
\eeq
where $g^2$ is the coupling constant for the sigma model.

For a generic sigma model in two spacetime
dimensions, both the field, $\boldl$, and the coupling constant, $g^2$,
are dimensionless.  It is well known that the coupling $g^2$ is
asymptotically free in two space-time dimensions \cite{zinn}.

The model can be generalized by relaxing the global symmetry to one
of $SU(N)$, taking $V = U^\dagger$.  The common assumption is to
impose a further constraint on the trace of $\boldl$.   
Then the only possible action remains as in (\ref{nonlinear}), but the
symmetry changes, as the model now defines a symmetric
space \cite{zinn,rdp_symmetric,symmetric}.  
For example, if $N=2M$ is even,
and one imposes the constraint that $\boldl$ is traceless,
$\tr \boldl = 0$, the symmetry becomes
$SU(2M)/S(U(M) \times U(M))$ \cite{zinn,rdp_symmetric,symmetric}.

To study the deconfining phase transition in $d+1$ spacetime dimensions,
one can construct an effective theory of straight, thermal Wilson
lines in $d$ spatial dimensions \cite{rdp}.  This effective theory
is valid over (spatial) distances $\vex \gg 1/T$, where $T$
is the temperature.  As a theory in two spatial dimensions, then,
the model we study is relevant to the transition in $2+1$ dimensions.
We will discuss the salient properties of this
effective theory as we go along.  For now, we only need to know that
in the effective theory, the Wilson line, which we denote as $\boldl(x)$,
is invariant under local gauge transformations, $\Omega(\vex)$, and global
\zn transformations,
\beq
\boldl(\vex) \rightarrow 
e^{2 \pi i/N} \;
\Omega(\vex) \; \boldl(\vex) \; \Omega^\dagger(\vex) \; .
\eeq

To accommodate the local gauge invariance, it is necessary
to include a $SU(N)$ gauge field, $A_i$.  Define the adjoint covariant
derivative as $D_i = \partial_i + i G [A_i,]$, 
where $G$ is the gauge coupling.  The action for $\boldl$ is then 
\beq
\tr \left| D_i \boldl \right|^2 \; ,
\label{nonlinear_gauge}
\eeq
plus the usual action for the gauge field.  In the effective
theory for deconfinement, the $A_i$ represent the gauge potentials for
the (static) magnetic field.  

Unlike symmetric spaces, the trace of $\boldl$
cannot be constrained.  In fact, the simplest trace,
\beq
\ell = \frac{1}{N} \; \tr \boldl \; ,
\label{fundamental_loop}
\eeq
is an order parameter for deconfinement, the Polyakov loop
in the fundamental representation.  A nonzero value for the fundamental
loop signals that the vacuum spontaneously breaks the
global \zn symmetry in the deconfined phase.

As we cannot impose a constraint on the trace of $\boldl$, the
possible Lagrangians are far more complicated than those of the
usual nonlinear sigma model.  
To start with, there are terms with no derivatives; that is,
there is a potential for $\boldl$.  This is 
a sum over loops \cite{dlp},
\beq
{\cal V} = \sum_{\cals \epsilon \calrn} 
\; \kappa_\cals \; \real \; \ell_\cals \; .
\label{general_potential}
\eeq
Here, $\ell_\cals$ denotes the trace of a loop in the
representation $\cals$.  For example, the loop in the adjoint
representation is
\beq
\ellad = \frac{1}{N^2 - 1} \left( \, |\tr \boldl|^2 - 1 \right) \; .
\label{adjoint_loop}
\eeq
We always divide a loop by the dimensionality of the representation.
In this way, in the perturbative limit, when $\boldl = \bold 1$, 
all loops are normalized to be one.  

We assume that the breaking of the global \zn symmetry is spontaneous,
so that the only loops which contribute to the loop potential
are \zn neutral.  In the loop potential,
$\calrn$ denotes all possible \zn neutral representations.
The series starts with the adjoint loop.
Using the character expansion,
we only need take linear powers of loops, albeit in arbitrary representations.

Ref. \cite{rdp} discusses the construction of an effective theory in
$3+1$ dimensions, but at least formally, it is immediate to extend it
to $2+1$ dimensions.  
In any number of dimensions, the Wilson line is dimensionless.  
For the original gauge theory in $2+1$ dimensions, 
the gauge coupling squared, $g^2_3$, has dimensions of mass.
In the effective theory, 
classically the term for the electric field is identical to (\ref{nonlinear}),
with the sigma model coupling, $g^2 = g^2_3/T$ \cite{rdp}.
The gauge coupling in the effective theory is
$G^2 = g_3^2 T$, and so $G^2$ has dimensions of mass squared.
At one loop order, some potential terms, as in (\ref{general_potential}),
are induced, with $\kappa_\cals \sim T^2$.  Further terms are
induced by corrections to higher order;
these are then a power series in $g^2_3/T$, times $T^2$.
Besides terms with no and two derivatives, terms with
four and more derivatives also arise in the effective theory 
\cite{rdp}.  We stress that the construction of this effective theory
is only valid in perturbation theory,
where $g^2_3/T$ is small; implicitly, it is assumed that this is a reasonable
approximation, even down to the transition temperature.  While this
appears to be true in $3+1$ dimensions \cite{coupling2}, 
this has not yet been studied
in lattice simulations in $2+1$ dimensions \cite{3Dlattice}.

The dominant behavior of the theory in the infrared and ultraviolet
limits can be read off by the dimensionality of the couplings.
In the infrared limit, as all $\kappa_\cals$ have dimensions of
mass, they should dominate.  There is an important, if familiar
qualification: as a theory in two dimensions, fluctuations can be
important when the fields are light.  

Conversely, in the ultraviolet limit, terms with two derivatives dominate
over those with none.  Now of course, since we are ultimately interested
in an effective theory, valid only over large distances,
studying the ultraviolet behavior is something of 
an academic exercise.  It was our original hope that these theories might
be asymptotically free in {\it all} couplings.  If so, then as with the
usual nonlinear sigma model, one could be able to study the 
phase transition in the infrared limmit, in three dimensions,
by developing an expansion in $2 + \epsilon$ dimensions, $\epsilon > 0$.
In the next section and the remainder of the paper,
we consider the simplest extension of the nonlinear sigma model,
which involves the three couplings which contribute to quartic
interactions in the perturbative limit, Sec. \ref{magnitude_couplings}.
While that leaves an infinity of other couplings, since we find that
these three couplings are not uniformly asymptotically free, 
Sec. \ref{beta_functions_section}, there is no point in going any further.

With this {\it mea culpa} aside, we classify all terms with two
derivatives \cite{dlp}.
The most obvious is the original Lagrangian, times any \zn neutral loop:
\beq
\tr \left| D_i \boldl \right|^2 \; 
\real \; \ell_\cals  \;\;\; , \;\;\; e_\cals = 0\; .
\label{Lagrangian_one}
\eeq
$\real$ denotes the real part, while $e_\cals$ is the \zn charge of
a loop in the representation $\cals$.  As \zn charges, $e_\cals$ is
only defined modulo $N$.
For the time being, we do not bother with denoting coupling
constants, since we are only concerned with the types of terms
which can arise in the action.  

One can add more powers of $\boldl$ inside the trace in 
(\ref{Lagrangian_one}),
\beq
\real \; \tr \left( \left| D_i \boldl \right|^2 \boldl^p \right) \; 
\ell_\cals  \;\;\; , \;\;\; e_\cals = -p \; .
\label{Lagrangian_two}
\eeq

The global symmetry is only one of \zn, and not $U(1)$.  This implies
that instead of taking the trace of the complex conjugate square of
the covariant derivative, 
as in (\ref{Lagrangian_one}),
one can take the trace just of the square of the covariant derivative:
\beq
\real \; \tr \; (D_i \boldl)^2 \; 
\; \ell_\cals \;\;\; , \;\;\; 
e_\cals = -2 \; ;
\label{Lagrangian_three}
\eeq
this is then multiplied by a loop of charge $-2$ to ensure \zn invariance.
Further powers of $\boldl$ can be added inside the trace:
\beq
\real \; \tr \left( \left( D_i \boldl \right)^2 \boldl^p \right) \; 
\ell_\cals  \;\;\; , \;\;\; e_\cals = -p-2 \; .
\label{Lagrangian_four}
\eeq

This list continues, until one is only left with derivatives of loops:
\beq
\real \; \left( \partial_i \ell_\cals \right) 
\left( \partial_i \ell_{\cals'} \right) \; \ell_{\cals''}
\;\;\; , \;\;\; e_{\cals} + e_{\cals'} + e_{\cals''} = 0 \; .
\label{Lagrangian_five}
\eeq

This list is not meant to be exhaustive.  Clearly there is an infinite
set of relevant couplings.  As we shall see in the next section for
$N=2$, not all of these couplings are necessarily independent.
While this infinite set of couplings is much more complicated than
ordinary sigma models 
\cite{zinn,rdp_symmetric,mckane_stone,friedan,principal_chiral,symmetric,friedan},
it is dictated by the physics, and is unavoidable.

For models with $SU(N) \times SU(N)$ symmetry, and their associated symmetric
spaces, the exact $\cal S$-matrix can be determined by Bethe ansatz
\cite{principal_chiral}.  Because of the plethora of couplings, we do not
expect that our matrix model is soluble by similar means.

%%%%%%%%%%%%
\section{N=2 Matrix Models}
\label{sectiontwo}
%%%%%%%%%%%%

%%%%%%%%%%%%
\subsection{Classifying Lagrangians}
\label{classifying_Lagrangians}
%%%%%%%%%%%%

We turn to the case of two colors.  
Two colors is illuminating, since one can explicitly compute all
elements of the Lie group using the exponential parametrization:
\begin{eqnarray}
\boldl &=& \exp\left(i \; \vecal \cdot \vec{t}\;\right) \; , \\
&=& \cos\left(\fralt \right) \;
+\; 2 \; i \; \wha \cdot \vec{t} \; \sin\left(\frac{\alpha}{2}\right) 
\nonumber\; .
\end{eqnarray}
We use generators $t^a$, normalizing $\tr (t^a t^b) = \delta^{a b}/2$
(so $t^a = \sigma^a/2$, where the $\sigma^a$ are the
Pauli matrices).  We also denote
\beq
\vecal = \alpha \, \wha \;\;\; , \;\;\; \wha^2 = 1 \; .
\eeq

The Lagrangian of the gauged nonlinear sigma model is
\beq
\calla(\boldl) = 2 \; \tr \left| D_i \boldl \right|^2 \; .
\label{lagAa}
\eeq
We defer the definition of coupling constants until later.
For $N=2$, the covariant derivative is
\beq
D_i \, \wha = \partial_i \, \wha + 
G \, \vec{A}_i \times \wha \; .
\label{cov_der}
\eeq
The gauge field only couples to the isovector, $\wha^a$, and not to the
isoscalar, $\alpha$.  For $N=2$, the Lagrangian of the sigma model is then
\beq
\calla(\vecal) = 
\left(\partial_i \, \alpha \right)^2
+ 4 \; \sin^2 \left(\fralt \right) \left(D_i \wha \, \right)^2  \; .
\label{lagAb}
\eeq
To avoid clutter, the dependence of $\calla(\vecal)$ on
the gauge field, $\vec{A}_i$, is left implicit.

The next term, as in the series in (\ref{Lagrangian_two}), is to
multiply the adjoint loop times this term.  We subtract off the value
of the loop when $\boldl = \bold 1$, and so define
\begin{eqnarray}
\callb(\boldl) &=& \frac{3}{2} \;
\left(1 - \ell_{ad} \right) \; 
\tr \left| D_i \boldl \right|^2  \; , \label{lagBa}\\
& = &
\sin^2\left( \fralt \right) \calla(\vecal) \; .
\label{lagBb}
\end{eqnarray}

Another possible kinetic term involves the loop in the fundamental,
or doublet, representation:
\begin{eqnarray}
\callc(\boldl) &=& 4 \left( \partial_i \ell \right)^2 
\label{lagCa} \; , \\
&=& \sin^2\left( \fralt \right) \left( \partial_i \alpha \right)^2 \; .
\label{lagCb}
\end{eqnarray}
This is a term as in (\ref{Lagrangian_five}).  As this is formed
exclusively from the loop, it depends only upon the magnitude of
$\vecal$, and not upon its direction in isospin space, $\wha$.  
For $SU(2)$, $\ell = \tr \boldl/2$ is automatically real, so it doesn't
matter if one takes just the square of the derivative of $\ell$, or
the complex conjugate square.  Lastly, as the loop is gauge invariant,
$\callc$ is independent of the gauge field.

A term like that in (\ref{Lagrangian_four}) is
\beq
\tr \left( D_i \boldl \right)^2 
= - \calla(\vecal) + \callc(\vecal) \; .
\label{lagD}
\eeq
This identity is special to $N=2$, and shows that 
this is not a new, independent coupling.

It is clear that there is an infinity of possible couplings.
In going from the original Lagrangian, (\ref{lagAa}), to (\ref{lagBa}),
we multiplied by the adjoint loop minus one, which for $N=2$ is
proportional to 
$\sin^2(\alpha/2)$.  We can continue this process 
to infinite order, multiplying 
$\calla$ and $\callc$ by higher and higher
powers of $\sin^2(\alpha/2)$.  All of these
are independent couplings, with dimensionless coupling constants.  

For $N=2$, the Lagrangian $\calla$ is invariant under an extended global
symmetry of $(SU(2) \times SU(2))/Z(2)$, which is isomorphic
to $O(4)$ \cite{mckane_stone}.  
It is instructive to write the Lagrangians in terms of $O(4)$ fields.
This is done by introducing 
\beq
\sigma = \cos(\alpha/2),  \;\;\;\;\vec{\pi}
= \sin(\alpha/2) \wha\;,
\eeq
and then forming the $O(4)$ vector $\vec{\phi} = (\sigma,\vec{\pi})$.
The matrix constraint then becomes 
$\vec{\phi}\,^2 = \sigma^2 + \vec{\pi}\,^2 = 1$.
Dropping the gauge fields, we find
\beq
\calla(\vec{\phi}) = 4 \; ( \partial_i \vec{\phi} \, )^2
= 4 \left( \left(\partial_i \vec{\pi} \right)^2 + 
\frac{(\vec{\pi} \cdot \partial_i \vec{\pi})^2}{1 - \vec{\pi}^2} \right) \; .
\eeq
The doublet loop $\ell = \sigma$, so the new couplings in our matrix model are
\beq
\callb(\vec{\phi}) = ( 1 - \sigma^2) ( \partial_i \vec{\phi} \,)^2 
= \vec{\pi}^2 \left( \left(\partial_i \vec{\pi} \right)^2 + 
\frac{(\vec{\pi} \cdot \partial_i \vec{\pi})^2}{1 - \vec{\pi}^2} \right) \; ,
\eeq
and
\beq
\callc(\vec{\phi}) = 4 \left( \partial_i \sigma \right)^2 
= 4 \; \frac{(\vec{\pi} \cdot \partial_i \vec{\pi})^2}{1 - \vec{\pi}^2}  \; .
\eeq
It is clear that only $\calla$ is $O(4)$ invariant.  We write
$\callb$ and $\callc$ using both $O(4)$ degrees
of freedom, and in terms of the $\vec{\pi}$ field.
We do this because 
while the independent degrees of freedom are the $\vec{\pi}$ fields, it
is instructive to see how the $O(4)$ symmetry is broken in the new couplings.
In terms of the $\vec{\pi}$'s, (\ref{lagD})  is especially simple,
\beq
\tr \left( D_i \boldl \right)^2 
= - \; 4 \; \left(\partial_i \vec{\pi} \right)^2\;,
\eeq
just a free kinetic term for the pions.

The $\beta$-functions can be computed using the $\vec{\pi}$'s,
but we find the $\vec{\alpha}$'s more convenient.

%%%%%%%%%%%%
\subsection{Coupling Constants}
\label{magnitude_couplings}
%%%%%%%%%%%%

We introduce the coupling constants $g^2$, $\xi$, and $\lambda$ in
the bare Lagrangian as
\beq
\call = \frac{1}{2 g^2} \; \calla + \frac{\xi}{g^4} \; \callb
+ \frac{\lambda}{g^4} \; \callc \; .
\label{Lagrangian_couplings}
\eeq
To understand why we introduce the coupling constants, 
$\xi$ and $\lambda$, with overall factors of $1/g^4$, we first note that
\beq
\calla(\alpha) = 
\frac{4}{\alpha^2} \sinsqalh
\left( (D_i \vecal)^2 - (\partial_i \alpha)^2 \right) +
(\partial_i \alpha)^2 \; .
\eeq
This form is useful in the perturbative limit.  The simplest perturbation
theory is to expand about 
zero field, taking $\vecal \Rightarrow g \vec{\beta}$.
Expanding to quartic order in $\beta$, 
\beq
\frac{1}{2g^2} \calla = \frac{1}{2} ( D_i \vec{\beta}\, )^2 
- \frac{g^2}{12} \; \vebesq ( D_i \vec{\beta} \, )^2  \; 
+ \ldots \; ,
\label{pert_trivial_A}
\eeq
\beq
\frac{\xi}{g^4} \callb \approx \frac{\xi}{4} \; \vebesq
( D_i \vec{\beta} \, )^2 \; + \ldots \; , 
\label{pert_trivial_B}
\eeq
and
\beq
\frac{\lambda}{g^4} \; \callc \approx
\frac{\lambda}{4} 
( \vec{\beta} \cdot \partial_i \vec{\beta} \, )^2 \; + \ldots \; .
\label{pert_trivial_C}
\eeq
Thus $\calla$ includes the free part of the Lagrangian, plus quartic
interactions $\sim g^2$.  For zero background field,
the new terms in the Lagrangian, $\callb$ and $\callc$, do not contribute 
to the quadratic part of the Lagrangian: they only contribute to
quartic interactions $\sim \xi$ and $\sim \lambda$, respectively.  
Hence the couplings $g^2$, $\xi$, and $\lambda$ 
are all couplings which should be included to one loop order.

This continues with the terms neglected above: a term such
as $(\sin(\beta/2))^{2 n} \calla$ first contributes to a
$2n + 2$ point function of the $\vec{\beta}$'s.  Thus the proper
normalization of such a term is a new coupling constant times
$1/g^{2(n+1)}$.  

This explains why we concentrate on the above three couplings: they
are uniquely the only couplings which contribute to quartic interactions
in the perturbative limit.  

%%%%%%%%%%%%
\section{Background Field Method}
\label{bfm}
%%%%%%%%%%%%

%%%%%%%%%%%%
\subsection{Possible Classical Fields}
\label{pcf}
%%%%%%%%%%%%

We wish to compute the renormalized Lagrangian to one loop order.
As is well known, to do so it is easiest to use the background field method
\cite{zinn}.  We take some classical, background field $\vecal$,
and expand in a quantum field $\vec{\beta}$.  To one loop order,
it is only necessary to expand 
to quadratic order in the quantum fluctuations, $\vec{\beta}$.
The price paid is that all dependence upon the background field must be kept.
Thus to ease our labor, we want to choose the simplest possible
background field we can.  

The background field cannot be too simple, however.  To one loop order,
various counterterms are generated.  We need a background field which
allows us to distinguish which counterterms contribute to the renormalization
of which terms in the bare Lagrangian.

Consequently, even in principle, it does not suffice to expand about a trivial
background: 
while the quartic interactions in $\callc$ differ,
(\ref{pert_trivial_C}), those in $\calla$ and $\callb$,
(\ref{pert_trivial_A}) and (\ref{pert_trivial_B}), are the same.

Thus we need to expand about a background which is not trivial.
The next simplest possibility is to take a field which lies
everywhere in the same direction in isospin space, so that
$\partial_i \widehat{\alpha} = 0$.  For such a field, the Lagrangian becomes
\beq
{\cal L} = \frac{1}{g^2}
\left( 1 + \frac{(\xi + \lambda)}{g^2} 
\sinsqalh \right) \; (\partial_i \alpha)^2\; .
\label{abelian_ansatz}
\eeq
This background field allows us to separate contributions
to $\calla$ from those to $\callb$ and $\callc$, but we can't disentangle
which terms contribute to $\callb$, and which to $\callc$.  
In terms of $\beta$-functions, we could determine that for
$g^2$, and the sum of $\xi + \lambda$, but not for $\xi$ and $\lambda$
by themselves.

Another possibility is to expand for a fixed direction in
isospin space, but in the presence of a background gauge
field, using the covariant derivative, $[A_i,\widehat{\alpha}]$, 
to separate the various terms.  We did not do this, 
because in order to respect gauge invariance, it is
necessary to include fluctuations in the gauge field.
(This was checked by explicit calculation.)
Including the fluctuations in the gauge
field seems unduly complicated, since they don't contribute to the
ultraviolet limit in two dimensions; their presence would be merely
as a bookkeeping device, to sort out the different terms in the renormalized
Lagrangian.

The final alternative is to expand about a completely general background
field, whose isospin direction changes in space,
$\partial_i \widehat\alpha \neq 0$.  
For an arbitrary field $\vecal$, it is clear that
$\calla(\vecal)$, $\callb(\vecal)$, and $\callc(\vecal)$ 
represent distinct Lagrangians, 
(\ref{lagAb}), (\ref{lagBb}), and (\ref{lagCb}).
Thus we can certainly pick out the contributions to different
terms in the Lagrangian which are generated at one loop order.

Indeed, the necessity of distinguishing between different terms in the 
renormalized Lagrangian is why we limit our calculations to two colors.
The three couplings which we consider for $N=2$ exist for $N \geq 3$,
given in (\ref{lagAa}), (\ref{lagBa}), and (\ref{lagCa}).
Taking the classical field as
$\boldl_{cl} = \exp(i \alpha^a t^a)$, with $t^a$ the generators of
$SU(N)$, it is natural to take the $\alpha^a$ to lie in the Cartan
subalgebra.  Indeed, the simplest form is to take $\alpha^a$ proportional
to a single generator, related to global $Z(N)$ transformations,
\beq
\alpha^a = \alpha \; t_N \;\;\; , \;\;\;
t_N = \left(
\begin{array}{cc}
{\bf 1}_{N-1} & 0      \\
0             & -(N-1) \\
\end{array}
\right)
\; .
\label{N3_ansatz}
\eeq
This is precisely like the $SU(2)$ ansatz where 
$\partial_i \widehat{\alpha} = 0$, 
(\ref{abelian_ansatz}), since in $SU(2)$ we can
always chose the fixed direction in isospin space to lie along $t^3$.
As for $N=2$, when $N \geq 3$
the ansatz of (\ref{N3_ansatz}) is not sufficient to
distinguish between $\xi$ and $\lambda$, with the sum of
$\calla + \callb + \callc$ like that of (\ref{abelian_ansatz}).  
Thus to determine the separate $\beta$-functions for $\xi$ and $\lambda$,
it is necessary to expand in a field which lies in two
distinct directions.  Since there are $N-1$ elements of the Cartan subalgebra,
when $N \geq 3$ we can chose the $\alpha^a$ to lie in
two, different elements of the Cartan subalgebra.  
Thus extending our calculation
from $N=2$, to $N \geq 3$, is straightforward to do.  It is not difficult to
see, however, that this will be tedious; even the form of the
classical field is messy.  In Sec. \ref{conclusions},
we comment on results for $N \geq 3$, using the ansatz of (\ref{N3_ansatz}).
We use this to make an obvious guess about
the leading form of the $\beta$-functions when $N \geq 3$.  

%%%%%%%%%%%%
\subsection{Explicit Expansion}
%%%%%%%%%%%%

We expand
\beq
\boldl = \boldl_{cl} \; \boldl_{qu} =
\exp(i \, \vecal \cdot \vec{t}\; ) \exp(i g \,\vec{\beta} \cdot \vec{t}\;) \; ,
\label{bckd_field_def}
\eeq
where $\vecal$ is the classical, background field, and 
$g \vec{\beta}$ the quantum field.  The factor of $g$ is introduced to
simplify later results, as in Sec. \ref{magnitude_couplings}.
This ansatz is convenient in computing
$\calla(\boldl)$ and $\callb(\boldl)$, since
\begin{eqnarray} 
\tr \left( \partial_i \boldl^\dagger \, \partial_i \boldl  \right) &=& 
\tr  \!\! \left( \right.
\partial_i \boldl^\dagger_{cl} \, \partial_i \boldl_{cl} 
\label{expansion_traceA}\\
& + & 2 \; \boldl_{qu} \, \partial_i \boldl^\dagger_{qu} \;
\boldl^\dagger_{cl} \, \partial_i \boldl_{cl}  
+ \partial_i \boldl^\dagger_{qu} \; \partial_i \boldl_{qu} \left.
\right)\nonumber \; . 
\end{eqnarray}
Instead of the ansatz in (\ref{bckd_field_def}), we could have taken 
$\boldl = \exp(i (\vecal + g \vec{\beta}) \cdot \vec{t})$.
While this appears simpler, the expansion of (\ref{expansion_traceA})
is then much more complicated.

The product 
\beq
\boldl_{cl}^\dagger \; \partial_i \boldl_{cl} \; = 
i \; \vec{\cala}_i \cdot \vec{t} \; ,
\eeq
is like a gauge field, albeit one formed from a pure gauge transformation,
generated by the field $\boldl_{cl}$.  Explicitly,
\beq
\vec{\cala}_i = \partial_i \alpha \; \widehat{\alpha}
+ \sin(\alpha) \; \partial_i \widehat{\alpha}
+ 2 \, \sinsqalh \; \widehat{\alpha} 
\times \partial_i \widehat{\alpha} \; ,
\eeq
where we have used the identity
\beq
\vecal \cdot \vec{t} \; \; \vec{\beta} \cdot \vec{t} =
\frac{1}{4} \; \vecal \cdot \vec{\beta} \; {\bf 1}
+ \frac{i}{2} \; \vec{t} \cdot (\vecal \times \vec{\beta}) \; ;
\eeq
$\vec{t} \cdot (\vecal \times \vec{\beta}) = \epsilon^{a b c}
t^a \alpha^b \beta^c$, where $\epsilon^{a b c}$ is the completely
antisymmetric tensor.  Further,
\beq
\boldl_{qu} \, \partial_i \boldl_{qu}^\dagger
\approx - i \; \partial_i \vec{\beta} \cdot \vec{t}
+ \frac{i}{2} \;
\vec{t} \cdot (\vec{\beta} \times \partial_i \vec{\beta}\,)
+ \ldots 
\eeq
To quadratic order, the doublet loop is
\begin{eqnarray}\nonumber
\ell = \frac{1}{2} \tr \boldl 
&=& \cosalh - \frac{g}{2} \, \sinalh \,
\widehat{\alpha} \cdot \vec{\beta}\\
&-& \frac{g^2}{8} \, \cosalh \, \vebesq + \ldots
\end{eqnarray}

Expanding the entire Lagrangian
to linear order in the quantum fluctuations, and integrating derivatives
by parts so that none acts on $\vec{\beta}$, we find
\begin{eqnarray}
\delta^1 \call &=& \frac{\vec{\beta}}{g} \cdot \left(
- \, \partial_i \vec{\cala}_i \label{delta_call_1} \right.\\
&+& \tillam \, \sinalh \left(\partial^2 \cosalh \right) \wha 
\nonumber\\
&+& \left. \frac{\tilxi}{2}
\left( \calla(\vecal) \sin(\alpha) \; \wha
- 4 \; \partial_i \left( \sinsqalh \vec{\cala}_i \right) \right) \right) \; .
\nonumber
\end{eqnarray}
We introduce the modified couplings
\beq
\tilxi = \frac{\xi}{g^2} \;\;\; , \;\;\;
\tillam = \frac{\lambda}{g^2} \; ,
\label{modified_couplings}
\eeq
as these arise naturally in the computation to one loop order.

The equation of motion is $\delta^1 \call = 0$.
That of the ordinary nonlinear sigma model is just
the first term on the left hand side 
in (\ref{delta_call_1}), $\partial_i \vec{\cala}_i = 0$.
The additional terms, $\sim \tilxi$ and $\tillam$,
are new contributions to the equation of motion in this matrix model.

The computation of terms to quadratic order in $\vecbe$ is 
straightforward.  We integrate by parts freely, organizing terms so
that as few derivatives as possible act on $\vecbe$'s, and as many
on the background field, $\vecal$.  Doing so, we find that we
can organize terms according to the maximum number of derivatives which
act on $\vecbe$.  

Terms with up to two derivatives acting on $\vecbe$ are
\begin{eqnarray}
\delta^2_2 \call &=&
\frac{1}{2} \left( 1 + 2 \, \tilxi \, \sinsqalh \right)
(\partial_i \vec{\beta})^2 \nonumber\\
&+& \tillam \left( \partial_i \left( \sinalh \wha \cdot \vec{\beta} 
\; \right) \right)^2 \; \label{quadratic_twoderiv} .
\end{eqnarray}
In writing $\delta^2_2 \call$, the superscript 
denotes an expansion to quadratic order in $\vecbe$;
the subscript denotes the maximum number of derivatives acting on $\vecbe$.

Terms with one derivative of $\vecbe$ are
\begin{eqnarray}
\delta^2_1 \call &=&
-\frac{1}{2} \left( 1 + 2 \, \tilxi \, \sinsqalh \right)
\vec{\cala}_i \cdot (\vec{\beta} \times \partial_i \vec{\beta})
\nonumber\\
&+& \tilxi \, \sin\alpha \; ( \wha \cdot \vec{\beta}\,) \;
\vec{\cala}_i \cdot \partial_i \vec{\beta} \; .
\label{quadratic_onederiv} 
\end{eqnarray}

Lastly, terms with no derivatives with respect to $\vecbe$ are
\begin{eqnarray}
\delta^2_0 \call &=&
\tillam \; \cosalh \; \left( \partial^2\cosalh \right) \, \vebesq 
\label{quadratic_noderiv} \\
&+& \frac{\tilxi}{4} \; \calla(\vecal)
\left( \vebesq - 
\sinsqalh \left( \vebesq + ( \wha \cdot \vec{\beta})^2 \right) 
\right) \; .\nonumber
\end{eqnarray}

Because we have normalized the quantum field to include a factor
of $g$, terms to quadratic order which are independent of any couplings
arise from the Lagrangian of the usual nonlinear sigma model.
These are the first terms on the left hand side 
in (\ref{quadratic_twoderiv}) and (\ref{quadratic_onederiv}).
The first term on the left in (\ref{quadratic_twoderiv}) 
is just the usual free kinetic term, 
$(\partial_i \vecbe)^2/2$.  The first term on the left in
(\ref{quadratic_onederiv}),
$-\vec{\cala}_i \cdot (\vec{\beta} \times \partial_i \vec{\beta})/2$,
represents the interactions in the usual nonlinear sigma model.
All other terms, which are proportional to either $\tilxi$ or $\tillam$,
arise from the new couplings in a matrix model.

There is an important restriction on the coupling constants in these
models.  Of course we can only compute in weak coupling, when
$g^2$, $\lambda$, and $\xi$ are all $\ll 1$.  Since the
new couplings in a matrix model affect the kinetic term, though, 
so that the usual kinetic term dominates at high momentum,
it is necessary that the new terms have small couplings:
\beq
\tilxi \; , \; \tillam \ll 1 \;\;\; , \;\;\;
\xi \; , \; \lambda \ll g^2 \ll 1 \; .
\label{coupling_restriction}
\eeq

In fact this restriction arises naturally in the construction
of the effective theory \cite{rdp}.  As discussed following
(\ref{adjoint_loop}), the sigma model coupling
$g^2 = g^2_3/T$.  Corrections at one loop order at one loop
order are $g^2_3/T$ times this coupling; hence $\xi/g^2$
and $\lambda/g^2$ are $\sim g^2_3/T$, or
$\xi,\lambda \sim (g^2_3/T)^2$.  

Considered purely as a theory in two dimensions, it is
possible to satisfy (\ref{coupling_restriction}), and still 
have contributions at one loop order 
dominate over those at higher loop order.
At one loop order, corrections are 
generically $\sim g^2$, $\lambda$, and $\xi$ times the quantity
at tree level.  At two loop order, corrections are then
$\sim g^4$, $g^2 \lambda$, $\lambda^2$, and so on, times the quantity
at tree level.  Thus it is possible to take --- for example ---
$\lambda$ and $\xi \sim g^3$.  In this case, (\ref{coupling_restriction})
is satisfied, since 
$\tilxi \sim \tillam \sim g \ll 1$, but terms at one
loop order, $\sim \lambda$ and $\xi$, 
are still larger than two loop terms from the ordinary
sigma model, $\sim g^4$.

%%%%%%%%%%%%
\section{One Loop Effective Action}
\label{one_loop}
%%%%%%%%%%%%

%%%%%%%%%%%%
\subsection{What to Compute}
%%%%%%%%%%%%

Having obtained the effective Lagrangian to quadratic order in $\vecbe$,
to one loop order the effective action involves the complete
inverse propagator in the presence of the background field:
\beq
\calseff = \frac{1}{2} \; \tr \log \left( 
\nabla^{a b} + \widetilde{\calv}_i^{a b} \partial_i
+ \widetilde{\calm}^{a b}\right) \; .
\label{effective_action1}
\eeq
The trace involves summation over isospin indices and space-time
momenta.  

In writing the inverse propagator, we take the liberty to
freely integrate terms by parts.  Terms in the inverse propagator
can be categorized according 
to the number of times the derivative operator appears.  Note that this
corresponds to a derivative on the quantum
field, $\vecbe$, and not to a derivative on the background field, $\vecal$.

The term with two derivatives is a Laplacian in the background field,
\begin{eqnarray}
\nabla^{a b} &=& -\; 
\partial^2
\; \delab \label{inverse_prop_twoderiv} \\
&+& 2 \;\tilxi \; \veclepar_i
\sinsqalh \vecripar_i \; \delab \nonumber\\
&+& 2 \; \tillam 
\left( \sinalh \wha^a \right) \; \veclepar_i
\; \vecripar_i 
\left( \sinalh \wha^b \right) \; .
\nonumber
\end{eqnarray}
The first term on the right hand side, 
$-\partial^2$, is the usual Laplacian for a free field.

The term free of the derivatives is a type of mass term,
\begin{eqnarray}
\widetilde{\calm}^{a b} &=&  2 \;\tillam \; \cosalh \; 
\partial^2 \!\cosalh \; \delab 
\label{inverse_prop_noderiv1}\\
&+& \frac{\tilxi}{2} \; \calla(\vecal) 
\left( \cossqalh \delab - \sinsqalh \alaalb \right) \; .
\nonumber
\end{eqnarray}

As for the second term in (\ref{effective_action1}), 
we first consider a toy Lagrangian.  Let a field $\beta^a$ interact
through a derivative interaction with a background field 
$\widetilde{\calv}_i^{a b}$:
\beq
\call_{\tilde{\cal V}} = \frac{1}{2} ( \partial_i \beta^a)^2
+ \widetilde{\calv}_i^{a b} \; \beta^a \, \partial_i \, \beta^b \; .
\label{lag_one_derivA}
\eeq
We do not assume anything about the symmetry of 
$\widetilde{\calv}_i^{a b}$ in the
isospin indices.  The Laplacian in this background field is
\beq
\nabla_{\tilde{\cal V}}^{a b} = - \; \partial^2 \; \delta^{a b}
+ \frac{1}{2} \left( \widetilde{\calv}_i^{a b} \vecripar_i
+ \veclepar_i \widetilde{\calv}_i^{b a} \right) \; .
\label{lag_one_derivB}
\eeq
The two terms in $\sim \widetilde{\calv}_i$ 
arise because the derivative can act either
on the $\vecbe$ which appears on the left, or that which appears on
the right.  For the last term, where the derivative acts to the left,
it is permissible to integrate by parts.  Doing so, the derivative
acts either as an operator, or it acts on the background field:
\beq
\nabla_{\tilde{\cal V}}^{a b} = - \; \partial^2 \; \delta^{a b}
+ \frac{1}{2} \left( \widetilde{\calv}_i^{a b} - 
\widetilde{\calv}_i^{b a} \right) \vecripar_i
- \frac{1}{2} \left(\partial_i \widetilde{\calv}_i^{b a} \right) \; .
\label{lag_one_derivC}
\eeq
This result is general, as we have made no assumption about the order
of the momenta.
The result in (\ref{lag_one_derivC}) can also be derived more carefully,
without any cavalier integration by parts. 
One expands the original
Lagrangian, (\ref{lag_one_derivA}), into terms which are symmetric
and anti-symmetric in the isospin indices.  This is done both
for the background field $\widetilde{\calv}_i^{a b}$,
and for $\beta^a \partial_i \beta^b$.  After some algebra, one
finds the same result as in (\ref{lag_one_derivC}). 

In the problem at hand, $\widetilde{\calv}_i^{a b}$ is given by
\begin{eqnarray}
\widetilde{\calv}_{i}^{a b}  &=& 
 - \frac{1}{2}\; \epsilon^{a b c} \left( 1 + 2 \; \tilxi \; \sinsqalh \right)
\cala_i^c   \nonumber\\
&+&  \tilxi \;
\sin(\alpha) \; \wha^a \, \cala_i^b\; .
\end{eqnarray}
Eq.\;(\ref{lag_one_derivC}) now contains one term that has 
one derivative on the 
quantum field, $\vec{\beta}$:
\begin{eqnarray}
\calv_{i}^{a b} \equiv \widetilde{\calv}_{i}^{ab} - 
\widetilde{\calv}_{i}^{ba}&=& 
 - \; \epsilon^{a b c} \left( 1 + 2 \; \tilxi \; \sinsqalh \right)
\cala_i^c   \nonumber\\
&+&  \tilxi \;
\sin(\alpha) \left( \wha^a \cala_i^b - \wha^b \cala_i^a \right) \; ,
\label{inverse_prop_onederiv}
\end{eqnarray}
and one term, $-\partial_i\;\widetilde{\calv}_i^{ba}$,  
which has no derivatives 
on $\beta$:
\beq
-\partial_i\;\widetilde{\calv}_i^{ba} =  - \;  \tilxi \; \partial_i
\left( \; \sin(\alpha) \; \wha^b \; \cala_i^a \right) \;.
\eeq
This is a type of mass term, and so we add it to that of
(\ref{inverse_prop_noderiv1}), and define 
\begin{eqnarray}
\calm^{a b} &=& - \;  \tilxi \; \partial_i
\left( \; \sin(\alpha) \; \wha^b \; \cala_i^a \right) 
\nonumber\\
&+& 2 \;\tillam \; \cosalh \; \partial^2 \!\cosalh \; \delab 
\label{inverse_prop_noderiv}\\
&+& \frac{\tilxi}{2} \; \calla(\vecal) 
\left( \cossqalh \delab - \sinsqalh \alaalb \right) \; .
\nonumber
\end{eqnarray}

The effective action is then
\beq
\calseff = \frac{1}{2} \; \tr \log \left( 
\nabla^{a b} + \calv_i^{a b} \vecripar_i 
+ \calm^{a b}\right) \; ,
\label{effective_action}
\eeq
where the term with one derivative now is like a coupling to a 
gauge field $\calv_i^{ab}$.

Computing the complete effective action for this inverse propagator would
be most involved.  However, our needs are much simpler.
We do not want all terms in the effective action, but
only the ultraviolet divergent terms which contribute to the
renormalization of the original Lagrangian.  
For this, 
we only need to expand the effective action to include terms with
two derivatives of the background field, 
$\vecal$.  There is an infinite series of
terms with higher numbers of derivatives of $\vecal$, but 
these are ultraviolet finite.  Terms with higher derivatives
do have infrared divergences, but these are cutoff by masses generated
by the loop potential.

The simplest term to compute is that from the mass term,
$\calm^{a b}$.  Like the original Lagrangian, this is already
of second order in derivatives of the background field, and
so can be expanded directly.  Terms involving the derivative operators,
$\calv_{i}^{a b} \partial_i$ 
and $\nabla^{a b}$, are more subtle to compute.
Some pieces of these operators are already of quadratic order in
derivatives of $\vecal$, but other pieces are not.  Thus we have to
take care, to ensure that we include all terms of quadratic
order in the background field.

Besides terms of quadratic order in derivatives
of the background field, there are terms of zeroth order in derivatives.
Such terms represent ultraviolet divergent renormalizations of the
loop potential.  In terms of the renormalization group, these contribute to
anomalous dimensions for the operators in the loop potential.  We 
do not consider such terms in the present work, since the first thing
to compute are the $\beta$-functions.

In the next three subsections, we discuss examples which illustrate how
to compute each of these terms.  We put all of the pieces together
in Sec. (\ref{beta_functions_section}).

We regularize integrals
by dimensional continuation from two to $2+\epsilon$ dimensions,
but checked that the same results are obtained with Pauli-Villars
regularization.  We also derived all results
using a mass to cutoff infrared divergences.  We do not present
the details of these checks, since they don't affect the final
results, but mention them to reassure the skeptical reader.

%%%%%%%%%%%%
\subsection{Terms with no derivatives}
\label{noderiv}
%%%%%%%%%%%%

The first example is trivial, the free energy of a mass term:
\beq
\frac{1}{2} \; \tr \log \left( - \partial^2 
+ m^2 \right) \; .
\eeq
Computing in $2+\epsilon$ dimensions, the ultraviolet divergent contribution
to the effective Lagrangian is
\beq
= \frac{m^2}{2} 
\; \int \frac{d^{2 + \epsilon}k}{(2 \pi)^{2+\epsilon}} \; \frac{1}{\overk^2} 
\approx - \; \frac{m^2}{4 \pi \epsilon} \;  .
\label{integral_noderiv}
\eeq
This result is valid only to $\sim 1/\epsilon$,
neglecting all contributions which are finite as $\epsilon \rightarrow 0$.
This integral is needed to compute
the contribution of $\calm^{a b}$ in (\ref{inverse_prop_noderiv}).

%%%%%%%%%%%%
\subsection{Terms with one derivative}
\label{onederiv}
%%%%%%%%%%%%
Since $\calv_i^{a b}$ is linear in derivatives
of the background field we have to expand it to 
quadratic order to obtain an ultraviolet divergent term of quadratic 
order in derivatives. 
This contributes to the effective action,
$\frac{1}{2} \, \tr \log \Delta^{a b}$, as
\begin{eqnarray}
&=& -\frac{i^2}{4} \calv_i^{a b} \;\calv_j^{a b} 
\int \frac{d^{2 + \epsilon}k}{(2\pi)^{2 + \epsilon}}
\; \frac{k^i k^j}{(\overk^{2})^2} \; , \nonumber\\
&\approx&
- \; \frac{1}{16 \pi \epsilon} \left(\calv_i^{a b}\right)^2
\; .
\label{quadratic_vs}
\end{eqnarray}
The factors of $i$ arise in going from coordinate to
momentum space, $\vecripar_i \Rightarrow i k^i$.

%%%%%%%%%%%%
\subsection{Terms with two derivatives}
\label{twoderiv}
%%%%%%%%%%%%

We turn to the expansion of terms containing up to two derivatives.

We start with the case where the background fields can be treated as constant.
Then the Laplacian
derived from (\ref{inverse_prop_twoderiv}) is just
the function
\begin{eqnarray}
\widetilde{\nabla}^{a b} &=&
\left( 1  +  2 \; \tilxi \; \sinsqalh \right) \delab 
\nonumber\\
&+& 2 \; \tillam \; \sinsqalh \; \wha^a \; \wha^b  \; ,
\label{inverse_prop_twoderiv_const} 
\end{eqnarray}
times the usual Laplacian for a massless field, $- \partial^2$.

For constant fields, it is easy computing the corresponding
propagator.  In fact, we can further simplify our algebra
by noting that in order to compute
the $\beta$-functions to one loop order, it suffices
to use the form of this propagator which is valid
only to linear order in $\tilxi$
and $\tillam$.  This propagator is
\begin{eqnarray}
\widetilde{\Delta}^{a b} &\approx& \left( 
\left( 1  -  2 \; \tilxi \; \sinsqalh \right) \delab \right.
\nonumber\\
&-& \left. 2 \; \tillam \; \sinsqalh \; \wha^a \; \wha^b  \right) \; ,
\label{prop_twoderiv_const} 
\end{eqnarray}
times the usual massless propagator, $1/(- \partial^2)$.

Notice that the restriction on the coupling constants in
(\ref{coupling_restriction}) is obvious in this form.  If
$\tilxi$ or $\tillam$ were not $\ll 1$, they would alter the usual
propagator of free field theory, and perturbation theory would be
more complicated.

As we explain in the next subsection,
it is necessary to use this modified propagator in computing the 
effects of the terms with no and one derivatives of the background
field.  

The problem is that in general, one is {\it not} allowed to treat
the background fields as constant.  To understand the problem,
consider the toy model,
\beq
\call_f = \frac{1}{2} (1 + f) \left( \partial_i \beta \right)^2 \; ,
\eeq
where $f$ is some background field.  We wish to compute the ultraviolet
divergences for terms which are quadratic in derivatives of $f$.

Besides such terms, there are also 
ultraviolet divergent terms which are constant in $f$; we ignore these
in the present model, as renormalizations of the loop
potential.  In two space-time dimensions, there are also
terms, quadratic in derivatives of $f$, which arise through the
conformal anomaly \cite{zinn}.  Terms from the conformal anomaly, though,
are ultraviolet finite, and so can be ignored in computing
$\beta$-functions.

The Laplacian for this toy model is
\beq
\nabla_f = - \; \partial^2 + \veclepar_i f \vecripar_i
\; .
\eeq
Terms of linear order in $f$ can be neglected, since they only
contribute to terms constant in $f$.  The first terms which depend
upon derivatives of $f$ arise at quadratic order.  We then
expand the effective action, $\frac{1}{2} \;\tr \log \nabla_f$, to
quadratic order.  We go to momentum space, with $p$ the momentum
going through the background field, $f$.  
Then the term of quadratic order in the
effective Lagrangian is
\beq
- \frac{1}{4} \; f(\overp\,) \left(\;
\int \frac{d^{2 + \epsilon}k}{(2 \pi)^{2+\epsilon}} \; 
\frac{( \overk\cdot(\overk+\overp\;))^2}{\overk^2(\overk+\overp\;)^2} 
\; \right) \; f(-\overp\,) \; .
\eeq
The structure of the
numerator is easy to understand.  This is a one loop diagram,
where the background field $f(\overp)$ couples to two quantum
fields.  The momenta at each vertex are $\overk$ and
$\overk + \overp$, with the coupling proportional to
the product of these momenta.

Since we only want the momentum dependent term, we can subtract off
the value of the integral for $\overp=0$.  This reduces a quadratically
divergent integral to one which is merely logarithmically divergent,
\beq
- \frac{1}{4} \; f(\overp\,)  \left(\;
\int \frac{d^{2 + \epsilon}k}{(2 \pi)^{2+\epsilon}} \; 
\frac{(\overk \cdot\overp\,)^2 
- \overk^2 \,\overp^{\;2}}{(\overk^2)^2} 
\; \right) \; f(-\overp\,) \; .
\eeq
Using the integral in (\ref{quadratic_vs}),
the contribution to the renormalized Lagrangian is
\beq
-\frac{1}{16 \pi \epsilon} \; \left( \partial_i f \right)^2 \; ,
\eeq
where we have reverted to coordinate space.

A slightly more involved 
toy Lagrangian, which most closely mimics our problem, is
\beq
\call_{f,h} = 
\frac{1}{2} (1 + f) \left( \partial_i \beta \right)^2 
+ \frac{1}{2} ( \partial_i  
(\vec{h} \cdot \vecbe ) )^2 \; .
\label{toy_f_h}
\eeq
By similar computation, 
the ultraviolet divergent counterterms at one loop order are
\beq
- \frac{1}{16 \pi \epsilon} \left(
(\partial_i f)^2 
+\; 4 \, (1 - f)\; ( \partial_i \vec{h}\,)^2
- 2 \;( \partial_i \vec{h}^2 )^2 \right) \; .
\label{result_f_h}
\eeq
%

%%%%%%%%%%%%
\section{$\beta$-functions to one loop order}
\label{beta_functions_section}
%%%%%%%%%%%%

With these results, we can compute the ultraviolet divergent
contributions to the Lagrangian at one loop order.

In doing so, it helps to recognize that this is all we want to do.
We only want terms which renormalize the bare Lagrangian, and
so only need keep terms $\sim \calla(\vecal)$, 
$\callb(\vecal)$, and $\callc(\vecal)$.  
At one loop order, we explicitly see that
new interactions, as discussed at the end of Sec. 
\ref{classifying_Lagrangians}, do appear.
These include, for example,
$\sin^4(\alpha/2) \calla(\vecal) \sim \sin^2(\alpha/2) \callb(\vecal) $,
and $\sin^2(\alpha/2) \callc(\vecal)$.  Such terms do have
ultraviolet divergences, which contribute to the $\beta$-functions
for these couplings.  We ignore these other couplings in the present work,
considering the three couplings which we do include as representative.
Partial results about the renormalization of such couplings are
given in Sec. \ref{conclusions}, (\ref{sum_couplings}).

Another important simplification is to recognize that terms at one
loop order are one power of the coupling constant times the bare Lagrangian,
(\ref{Lagrangian_couplings}).
Remembering the definitions of couplings in (\ref{modified_couplings}), 
terms $\sim g^2$ times those in the original Lagrangian generate
those $\sim 1$, $\tilxi$, and $\tillam$; terms
$\xi$ times the original Lagrangian generate contributions
$\sim \tilxi$, $\tilxi^{\,2}$, and $\tilxi \;\tillam$;
those $\lambda$ times the original Lagrangian generate
terms $\sim \tillam$, $\tilxi \;\tillam$, and $\tillam^{\, 2}$.  
Thus at one loop order, we have to include all terms $\sim 1$,
$\tilxi$, $\tillam$, $\tilxi^{\,2}$, $\tilxi \; \tillam$, and $\tillam^2$.

Of course, terms which are ultraviolet finite can be ignored completely.
This includes all terms with more than two derivatives of the background
field.

The easiest contribution to compute is that from terms which are already
of second order in derivatives, $\calm^{a b}$ in 
(\ref{inverse_prop_noderiv}).  Even so, we have to recognize that
we cannot use (\ref{integral_noderiv}) directly, but
must use the propagator in a background field.  Fortunately,
as we already have terms with two derivatives, we can use
the propagator for a constant, background field in (\ref{prop_twoderiv_const}):
\beq
- \; \frac{1}{4 \pi \epsilon} \; \widetilde{\Delta}^{a b}
\calm^{a b} \; ,
\label{constant_mass_term}
\eeq
which equals
\begin{eqnarray}
\frac{1}{\pi \epsilon}
\left( - \frac{3}{8} \; \tilxi \; \calla \right.
&+& \frac{1}{4}
\left( 2 \; \tilxi + 3 \; \tilxi \; \tillam
+ 3 \; \tilxi^{\,2}\right) \callb
\label{counterterms_no_derivs}\\
&+& \left. 
\frac{1}{8} 
\left(3 \; \tillam + 8 \; \tilxi^{\, 2} + 16 \; \tillam \; \tilxi
+ 4 \; \tillam^2 \right) \callc \right) \; .
\nonumber
\end{eqnarray}
Of course implicitly, 
all of the $\call$'s are functions of the background field, $\vecal$.

The next contribution is from terms which are of first order in
the derivatives, $\calv_{i}^{a b}$ in (\ref{inverse_prop_onederiv}).
Again, we cannot use the free propagator, but must use the
propagator in a constant, background field.  To one loop,
this contributes the ultraviolet divergent terms as in
(\ref{quadratic_vs}),
\beq 
- \; \frac{1}{16 \pi \epsilon}\;
\calv_{i}^{a b} \;\widetilde{\Delta}^{b c} \; 
\calv_{i}^{c d} \; \widetilde{\Delta}^{d a}
\; ,
\eeq
which is
\beq
\frac{1}{\pi \epsilon}
\left( \; \frac{1}{8} \; \calla 
+ \frac{1}{4} \; (\tillam - \; 2 \; \tilxi^{\, 2}\,) \; 
\left( \callc - \callb\right) \right)\; .
\label{counterterms_one_deriv}
\eeq

The final contribution is from terms with two derivatives.
Using the toy model of (\ref{toy_f_h}), where we identify
\beq
f = 2 \; \tilxi \;\sinsqalh \;\;\; , \;\;\;
\vec{h} = 2 \; \sin\left(\frac{\alpha}{2}\right) \;\wha \; ,
\eeq
then from (\ref{result_f_h}), we find that the ultraviolet
divergent contributions to the one loop renormalized Lagrangian is
\begin{eqnarray}
\frac{1}{\pi \epsilon} \left(
- \frac{1}{8} \; \tillam \; \calla \right.
&+& \frac{1}{4} \; \tilxi \; \tillam \; \callb 
\label{counterterms_two_derivs}\\
&+& \left. \frac{1}{8}
\left(  \tillam - 2 \; \tilxi^{\, 2} + 
4 \; \tillam^2 \right) \callc \right) \; .
\nonumber
\end{eqnarray}

The sum of the ultraviolet divergent counterterms at one loop order,
(\ref{counterterms_no_derivs}), (\ref{counterterms_one_deriv}),
and (\ref{counterterms_two_derivs}), is
\begin{eqnarray}
\call_{ct} &=& \frac{1}{\pi \epsilon} \left(
\frac{1}{8} \left(1  - 3 \; \tilxi - \tillam\right) \calla \right.
\label{counterterm_Lagrangian}\\
&+& \frac{1}{4} \left( 2 \; \tilxi -  \; \tillam
+ 4 \; \tilxi \; \tillam + 5 \; \tilxi^{\,2} \right) \callb \nonumber\\
&+& \left. \frac{1}{4} \left(
3 \; \tillam +  \tilxi^{\,2} + 8 \; \tilxi \; \tillam
+ 4 \tillam^2 \right) \callc \right) \; .
\nonumber
\end{eqnarray}

We write the renormalized Lagrangian as
\beq
\call_{ren} = \frac{1}{2  \calz_g \, g^2} \; \calla + 
\frac{\calz_\xi \, \xi}{\calz_g^2 \, g^4} \; \callb
+ \frac{\calz_\lambda \, \lambda}{\calz_g^2 \, g^4} \; \callc \; .
\label{renormalized_Lagrangian}
\eeq
The renormalization constants for the three couplings are $\calz_g$,
$\calz_\xi$, and $\calz_\lambda$.  In $2+\epsilon$ dimensions,
the couplings $g^2$, $\xi$, and $\lambda$ all have dimensions
of $\mu^\epsilon$, where $\mu$ is a renormalization mass scale.

The renormalization constants are fixed by requiring that
the ultraviolet divergences cancel in the sum of the counterterm
Lagrangian, $\call_{ct}$ in (\ref{counterterm_Lagrangian}), 
and the renormalized Lagrangian, $\call_{ren}$ in 
(\ref{renormalized_Lagrangian}) \cite{zinn}.
At one loop order, this determines the renormalization constants to be
\begin{eqnarray}
{\cal Z}_g &=& 1 + \frac{1}{4 \pi \epsilon} \; \left( g^2 - 3 \xi - 
\lambda \right)
\label{renormalization_constant_values} \; , \\
{\cal Z}_\xi &=& 1 + \frac{1}{4 \pi \epsilon} \; \left(
\frac{g^2 \lambda }{\xi} - 11 \; \xi - 6 \; \lambda  \right)
\nonumber \; , \\
{\cal Z}_\lambda &=& 1 + \frac{1}{4 \pi \epsilon} \; \left(
- \, g^2 - 14 \; \xi - 6 \; \lambda - \frac{\xi^2}{\lambda} \right) \; .
\nonumber
\end{eqnarray}
The inverse powers of the coupling constant appear worrisome; they
certainly do not arise in the $\beta$-functions of other nonlinear
sigma models \cite{zinn}.  

The $\beta$-function for the coupling $g^2$ is given by
\beq
\beta(g^2) = \epsilon \, g^2\left/\left( 1 - \epsilon \, 
\frac{d}{d \epsilon} \log 
\calz_g \right) \right. \; ,
\eeq
and similarly for the other couplings.

Using this definition, we find that the $\beta$-functions have a 
standard form:
\begin{eqnarray}
\beta(g^2) 
&=&  \epsilon \; g^2 + 
\frac{1}{4\pi} 
\left( - \; g^4 + 3 \; g^2\; \xi  + g^2 \; \lambda \right) + \ldots
\label{beta_functions} \; , \\
\beta(\xi) &=&  \epsilon \; \xi +
\frac{1}{4\pi} \left( - \; g^2 \; \lambda 
+ 11 \; \xi^2 + 6 \; \xi \; \lambda  \right) + \ldots
\nonumber \; , \\
\beta(\lambda) &=& \epsilon \; \lambda
+ \frac{1}{4\pi} \left( + \; g^2 \; \lambda + \xi^2
+ 14 \; \xi \; \lambda + 6 \; \lambda^2 \right) + \ldots \; .
\nonumber
\end{eqnarray}
\vspace{.01in}

In the limit where $\xi, \lambda \ll g^2$, the $\beta$-functions reduce to
\begin{eqnarray}
\beta(g^2) 
&\approx&  \epsilon \; g^2  
- \frac{1}{4\pi} \; g^4 + \ldots
\label{approximate_beta_functions} \; , \\
\beta(\xi) &\approx&  \epsilon \; \xi
- \frac{1}{4\pi} \; g^2 \; \lambda + \ldots
\nonumber \; , \\
\beta(\lambda) &\approx&  \; \epsilon \, \lambda
+ \frac{1}{4\pi}  \; g^2 \; \lambda + \ldots
\nonumber
\end{eqnarray}

These are the principal results of our paper.   
We now discuss the properties of these $\beta$-functions in two 
dimensions, where $\epsilon = 0$.

As discussed in Sec. \ref{classifying_Lagrangians}, 
the leading term in the $\beta$-function for the coupling $g^2$, 
$\beta(g^2) \sim - g^4/(4 \pi)$, is the same as for the $O(4)$ nonlinear
sigma model \cite{zinn,mckane_stone} (accounting for the somewhat
unconventional normalization of our coupling constant).
As we compute in the limit where $\xi$ and $\lambda$ are $\ll g^2$,
this term dominates those $\sim + g^2 \xi$ and $\sim + g^2 \lambda$ in
$\beta(g)$.  As the leading term is of negative sign, the coupling
$g^2$ is, inescapably, asymptotically free.

The diligent reader might wonder why we didn't save ourselves much
effort, and directly compute the
$\beta$-functions in the relevant limit, where $\xi$ and $\lambda$
$\ll g^2$.  In fact, this is what we did first.   We computed the
$\beta$-functions for a field which is constant in isospin space,
$\partial_i \wha = 0$, where it is much simpler to compute.
As seen from (\ref{abelian_ansatz}), however, this only gives the
$\beta$-function for the sum of the two couplings, 
$\xi + \lambda$.  After some labor, we found that this $\beta$-function
vanishes to the requisite order,
$\sim g^2 \xi $ and $\sim g^2 \lambda$.  From
(\ref{approximate_beta_functions}), we see that this is because
in $\beta(\xi)$ and $\beta(\lambda)$, the only terms of this order
are $\sim g^2 \lambda$.  These terms are equal and of opposite sign, 
so that up to terms of higher order, they cancel identically.

Thus in order to know the $\beta$-function for $\xi + \lambda$, it is
necessary to determine the nonleading terms in the $\beta$-functions, as in
(\ref{beta_functions}).
An important question is if it is possible to obtain an 
asymptotically free theory
in all three couplings.  
Notably, since we take both $\xi$ and $\lambda$
to be $\ll g^2$, the theory is sensible for either sign of each coupling.  
When $\xi \sim \lambda$, however, the theory is clearly not asymptotically
free: the dominant term in the $\beta$-functions for both $\xi$ and $\lambda$
is $\sim g^2 \lambda$.  But as each term appears with opposite
sign, (\ref{approximate_beta_functions}), whatever sign we take for $\lambda$,
one coupling is asymptotically free, and the other, infrared free.

Since $\lambda$ dominates the $\beta$-functions, one can also consider
the limit where we tune $\lambda$ to be much smaller
than $\xi$, with $\lambda \ll \xi \ll g^2$.  In this case, however, the full
calculation, (\ref{beta_functions}), show that the dominant term in
$\beta(\xi)$ and $\beta(\lambda)$ are each $\sim + \xi^2$.  Each of
these terms has a positive sign, and so produces infrared freedom, 
regardless of the sign of $\xi$.

%%%%%%%%%%%%
\section{Conclusions}
\label{conclusions}
%%%%%%%%%%%%

In this paper we computed the $\beta$-functions for the simplest
extension of a $O(4)$ nonlinear sigma model, which is a $N=2$ matrix model
with three coupling constants.  
To leading order in weak coupling, the result
is extremely simple, (\ref{approximate_beta_functions}).
In general, we find that there is no region of parameter space in
which all three coupling constants are asymptotically free,
(\ref{beta_functions}).

We conclude by discussing
unpublished work by one of us \cite{oswald}.  Both of these
calculations take a background field which lies along a single
direction in the Lie algebra.  As discussed in Sec. \ref{pcf}, this
ansatz is not adequate to give complete information about the
$\beta$-functions.  As we shall see, however, these limited
results indicate that when $N \geq 2$,
all matrix models include couplings which are {\it not} asymptotically free.

As discussed in Sec. \ref{variety}, there is an infinity of couplings which we
neglected.  Consider a background field which lies along one direction in
isospin space, $\partial_i \wha = 0$.  To represent the
neglected couplings, we add to the Lagrangian the term
\beq
\frac{\Xi_n}{(g^2)^{n+1}} \; \sin^{2 n}\left( \frac{\alpha}{2} \right) 
( \partial_i \alpha )^2 \; .
\label{sum_couplings}
\eeq
For $n=1$, this corresponds to (\ref{abelian_ansatz}), with
the coupling $\Xi_1 = \xi +\lambda$.  When $n \geq 2$, $\Xi_n$ 
is proportional to a sum of several, independent coupling constants.  
As in Sec. \ref{magnitude_couplings}, in 
expanding about a trivial background, with $\alpha \Rightarrow g \beta$, 
this term first contributes to a $2(n+1)$ point of $\beta$'s.  It is
for this reason that we normalize this term by an 
overall factor of $1/(g^2)^{n+1}$,
so that the $2(n+1)$ point function has coupling $\Xi_n$, independent of $g$.

The $\beta$-function for $\Xi_n$ is found to be 
$\sim + (n^2 - 1) g^2 \, \Xi_n$ \cite{oswald}.   This is valid to
$\sim g^2 \, \Xi_n$, and neglects subleading terms $\Xi_n^2$, {\it etc.}
This vanishes for $n=1$, in agreement with (\ref{approximate_beta_functions}).
When $n > 1$, though, the $\beta$-function for $\Xi_n$ is infrared free.  

Calculations were also done for $N \geq 3$.  
Following Sec. \ref{variety}, 
there are couplings, analogous to $g^2$, $\xi$, and $\lambda$, for all
$N \geq 3$.  Going through the same calculations as for $N=2$,
it is straightforward to determine the leading terms in the $\beta$-functions.
The leading term in the
$\beta$-function for $g^2$ is $\sim - N g^4$, so this coupling
is, of course, asymptotically free \cite{mckane_stone}.  
When $N \geq 3$, it is found that the 
$\beta$-function for $\xi + \lambda$
vanishes to leading order, $\sim g^2 \xi$ and $\sim g^2 \lambda$ \cite{oswald}.
This is precisely analogous to the results for $N=2$, 
and suggests that the form of (\ref{approximate_beta_functions})
is valid for arbitrary $N$, with 
$\beta(\xi) = - \beta(\lambda) \sim + g^2 \lambda$, up to
corrections $\sim \xi^2$, $\sim \xi \lambda$, and $\sim \lambda^2$.
That is, at least one of the two couplings is always infrared free.
We also expect that when $N \geq 3$, the infinite series of couplings to 
higher order, as in (\ref{sum_couplings}) for $N=2$, 
includes infrared free couplings.

In summary, these partial calculations strongly suggest that the
$\beta$-functions for $N \geq 3$ are very similar to those for $N=2$.
Generically, while the dominant coupling $g^2$ is asymptotically free,
there are always subdominant couplings which are infrared free.
We do not expect any region of parameter space in which all couplings
are asymptotically free.

We contrast this to the behavior of ordinary nonlinear sigma models
\cite{zinn,rdp_symmetric,mckane_stone,principal_chiral,symmetric}.  
These models only involve a single coupling constant,
which is asymptotically free in two dimensions.   In such theories, 
the symmetry is unbroken 
in two dimensions, and the theory is always in a massive phase.
Above two dimensions, an ultraviolet stable
fixed point appears, with $g^2 \sim \epsilon$.  The appearance of this
ultraviolet stable fixed point is assumed to be related to a phase transition
in which the symmetry breaks.
The properties of the transition, as determined from
this ultraviolet stable fixed point, working up 
in $2 + \epsilon$ dimensions, is believed to reflect the same
universal properties as for a linear sigma model, working down from
$4 - \epsilon$ dimensions \cite{zinn}.

There is a greater variety of phase transitions possible in matrix models,
which is
reflected in their Lagrangians.
Matrix models have potentials, and we can couple them to gauge fields.
In two dimensions, the terms in the potential, and the gauge coupling,
all have dimensions of mass squared, and so dominate in the infrared limit.

Conversely, since these terms have positive mass dimension, they do not
affect the ultraviolet behavior, in either
two or $2+\epsilon$ dimensions.  From present results, and those of
\cite{oswald}, it appears that the $\beta$-functions are
very similar for all $N \geq 2$.  

This is unlike the results of lattice simulations, which find that the order of
the deconfining transition changes as a function of $N$.  For theories in $2+1$
dimensions, the transition appears
to be of second order for $N = 2$, $3$, and possibly $4$, and of
first order for $N\geq 5$ \cite{3Dlattice}.  For theories
in $3+1$ dimensions, the
deconfining transition is of second order for $N=2$ \cite{su2}, 
and of first order for
$N \geq 3$ \cite{su3,teper,bringoltz,holland}.  

This demonstrates that the connection between the ultraviolet
behavior of $\beta$-functions in $2+\epsilon$ dimensions, 
and phase transitions in the infrared
limit, is not as immediate in matrix models, as it is in ordinary sigma models.
Certainly, before we can understand what happens nonperturbatively in
the infrared limit, we must first understand perturbation theory in
the ultraviolet limit.  This was the goal of the present work.  We hope that
it provides an impetus to better understand the rich structure of 
phase transitions 
possible in matrix models.

\section{Acknowledgments}

We thank P. Arnold, P. Fendley, and A. Rebhan for discussions.
R.D.P. also
thanks M. J. Bhaseen, F. Essler, R. Konik, and A. Tsvelik; 
lastly, J. Papavassiliou, with whom he first started working on this
problem.
The research of M.O. is supported by 
the U.S. Department of Energy grant DE-FG02-97ER41027; that of 
R.D.P., by grant DE-AC02-98CH10886.
R.D.P. also 
thanks the Alexander von Humboldt Foundation for their support.

\section{Appendix}

We use this opportunity to make a comment
about the correlation functions of 
Polyakov loops at large $N$.  Consider the two-point function of 
Polyakov loops,
\beq
\langle \tr \boldl(x)^\dagger \tr \boldl(0) \rangle
\sim \sum_i \; \exp(- m_i |x|) \;\; , \;\; x \rightarrow \infty ,
\label{two_point}
\eeq
where the summation is over all states which contribute to
this correlation function.  For a $SU(N)$ gauge theory without quarks,
the symmetry broken is $Z(N)$.  As $N\rightarrow \infty$,
this becomes $U(1)$.  By Goldstone's theorem, when $U(1)$ breaks,
there must be an associated massless particle.  At large but
finite $N$, there is a light, almost Goldstone particle;
as the mass of an almost Goldstone
particle is proportional to the parameter which breaks
the symmetry, which in the pure glue theory is $1/N^2$,
then $m^2_G \sim 1/N^2$, or $m_G \sim 1/N$.  
This almost Goldstone mode is separate from the
usual, perturbative excitation associated with the Debye mass.

This is trivial in three dimensions, but one might wonder if
it still holds in two dimensions, since then a continuous symmetry 
cannot break.
We suggest, however, that at large $N$, the two-point correlations
of (\ref{two_point}) 
will look similar in either two and three dimensions.  This is because
at large $N$,
all connected correlation functions are suppressed  by powers of $\sim 1/N^2$.
Thus it is possible to have a massless Goldstone boson at infinite $N$,
since then it is also non-interacting.

%\href{http://arXiv.org/}{[arXiv:]}

\end{document}